# FREQUENCY SELECTION FOR THE DIAGNOSTIC CHARACTERIZATION OF HUMAN BRAIN TUMOURS

CARLOS ARIZMENDI, ALFREDO VELLIDO, ENRIQUE ROMERO
Dept. de Llenguatges i Sistemes Informàtics (LSI).
Universitat Politècnica de Catalunya (UPC). 08034, Barcelona, Spain

**Abstract.** The diagnosis of brain tumours is an extremely sensitive and complex clinical task that must rely upon information gathered through non-invasive techniques. One such technique is magnetic resonance, in the modalities of imaging or spectroscopy. The latter provides plenty of metabolic information about the tumour tissue, but its high dimensionality makes resorting to pattern recognition techniques advisable. In this brief paper, an international database of brain tumours is analyzed resorting to an *ad hoc* spectral frequency selection procedure combined with nonlinear classification.

**Introduction**

Decision making in oncology is a sensitive undertaking, and even more so in the specific area of brain tumour oncology diagnosis, for which the costs of misdiagnosis are very high. In this area, in which data acquisition techniques are preferably indirect and non-invasive, clinicians should benefit from the use of at least partially automated decision support.

This study addresses the problem of human brain tumour diagnosis on the basis of biological signal data obtained by Magnetic Resonance Spectroscopy (MRS). MRS is a technique that can shed light on cases that remain ambiguous after clinical investigation. This technique has evolved rapidly over the past 15 years and has shown very encouraging correlations between brain tumour types and spectral patterns. *In vivo* MRS enables the quantification of metabolite concentrations non-invasively, thereby avoiding serious risks of brain damage. However, the introduction of MRS in clinical practice has been hampered by different problems: The first is associated with the acquisition of in vivo MRS signals from living tissues at magnetic fields low enough not to pose a threat to patients. Furthermore, clinicians often lack the training required to make sense of the MR spectral signal, limitation compounded by the fact that this task requires considerable experience from a radiologist. Additionally, the natural high dimensionality of the spectra, the presence of noise and artifacts, and the low amount of data available for specific pathologies (that is, for specific brain tumour types) complicates their diagnostic-oriented classification. Machine Learning (ML) and related data analysis methods can play a useful role in this setting.

We aim to investigate whether the contribution of adjacent variables/frequencies improves the class discrimination yielded by the use of individual, non-adjacent variables. To that end, we explore a new methodology that involves the combination of different techniques, namely the Moving Window and the between/within group variance analysis, for the reduction of the spectral dimensionality by selection of bandwidths based on the energy criteria. Nonlinear Artificial Neural Networks (ANN) will be used for classification in order to gauge the properties of the developed techniques.

|  | Number of cases | |
| --- | --- | --- |
| **Tumour class** | **SET** | **LET** |
| **a2**: Astrocytomas, grade II | 22 | 20 |
| **a3**: Astrocytomas, grade III. | 7 | 6 |
| **ab**: Brain abscesses | 8 | 8 |
| **gl**: Glioblastomas, giants cells | 86 | 78 |
| **hb**: Haemangioblastomas | 5 | 3 |
| **ly**: Lymphomas | 10 | 9 |
| **me**: Metastases | 38 | 31 |
| **mm**: Meningiomas grade I | 58 | 55 |
| **no**: Normal cerebral tissue, white matter | 22 | 15 |
| **oa**: Oligoastrocytomas grade II | 6 | 6 |
| **od**: Oligodendrogliomas grade II | 7 | 5 |
| **pi**: Pilocytic astrocytomas grade I. | 3 | 3 |
| **pn**: Primitive neuroectodermal tumours and medulloblastomas | 9 | 7 |
| **ra**: "Rare tumours" | 19 | 18 |
| **sc**: Schwannomas | 4 | 2 |

**Table 1**: Contents of the INTERPRET database. The table includes a list of the tumour types and the number of cases for each of them for both acquisition time echos.

## 1. Brain Tumours

The tumours of the central nervous system (CNS) represent around the 2% of the total of cancers diagnosed around the world. Annually, about 175,000 people are diagnosed with tumours that affect the CNS [1], out of which 29.000 in Europe [2]. The incidence ratio of this pathology is of 7 persons per 100,000. Different studies have shown that the distribution of the tumours by age is bimodal, with a peak in infants and another in adults between 40 and 70 [3]. A completely unambiguous diagnosis of a brain tumour, in terms of type and degree, can only be obtained by histopathological analysis of a brain biopsy.

*1.1. The Problem of Brain Tumour Classification Using MRS*

The current gold standard for classification of brain tumours is class labelling according to the World Health Organization (WHO) biopsy-based system. Biopsies require an invasive procedure with a risk of mortality of 0.2-0.8% and an estimate of morbidity in the range 2.4-3.5% [4, 5]. Additionally, only about a 91% of cases are truly identifiable through this test, which means that up to 9% of patients remain undiagnosed [6]. For these reasons, among others, it is essential to improve the classification of different types of brain tumours using non-invasive methods. Amongst these, we have MR Imaging (MRI), which provides accurate spatial resolution but little metabolic information, and Spectroscopy (MRS), which sacrifices spatial resolution to obtain a detailed metabolic fingerprint of the tumour tissue.

*1.2. INTERPRET: An International MRS Database of Human Brain Tumours*

This study relies on a database created under the framework of the European project INTERPRET, an international collaboration of centers from 4 different countries. The database includes single-voxel proton MRS (SV $^1$H-MRS), measured at long echo time (LET: 266 patients) and short echo time (SET: 304 patients). The total number of samples (spectral frequencies) was set to 512. For further details on data acquisition and processing, and on database characteristics, see, for instance, [7] and [8]. The database consists of the types of tumours listed in Table 1.

**2. Methods and Experiments**

*2.1. The Moving Window Technique*

The Moving Window (MW), is a technique in which a signal $X$ is multiplied by a window of magnitude 1 and constant width ($w$); a mathematical function or analytical processing is applied to this subset of samples, obtaining a result. Next, the window is displaced by one position, performing again the previous operation. The mathematical function can be used to find optimal sub-regions within informative frequency intervals, and also to perform extraction and/or selection of features. Figure 1 illustrates this idea. When $w = 1$, the MW moves along the signal, travelling from the first sample to the last, creating $m$-$w$+1 outlets ($m$ = number of samples). The width of window can vary from 1 to $m$, accounting for all possible adjacent sub-regions.

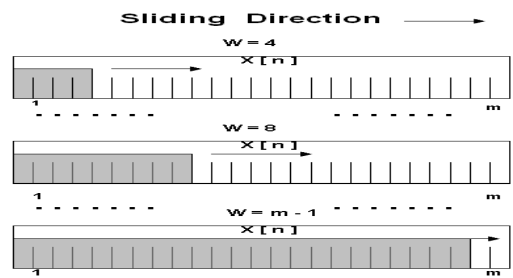

**Figure 1:** Schematic illustration of the Moving Window technique.

*2.2. MW and Between-Groups Variance Analysis*

We have developed a methodology based on the technique of MW and the analysis of between/within group variance, as a procedure for identifying those spectral frequencies (or intervals of continuous frequencies), with greater ability to discriminate between types of tumours. This is considered as a preliminary exploratory step towards a well-structured classification of tumour pathologies.

The technique of MW has been implemented in conjunction with the calculation of a standard ratio ($\lambda$) of between/within-group variance (in turn, *BGV* and *WGV*). The Euclidean metric was used for calculating the variances. The ratio is defined as:

$$\lambda = \frac{BGV}{WGV}$$

where

$$BGV = \sum_{i=1}^{n} \frac{\sqrt{\sum_{Q=1}^{W}(x_{i,Q} - \mu x_Q)^2}}{n\sqrt{w}} + \sum_{j=1}^{m} \frac{\sqrt{\sum_{Q=1}^{W}(y_{j,Q} - \mu y_Q)^2}}{m\sqrt{w}}$$

$$WGV = \frac{\sqrt{\sum_{Q=1}^{W}(\mu x_Q - \mu y_Q)^2}}{\sqrt{w}}$$

$x_{i,Q}$ and $y_{j,Q}$ are numerical vectors corresponding to patterns of *w*-dimensional characteristic, representing a number of variables and objects that conform the matrices **X** and **Y**, where $\mathbf{X},\mathbf{Y} \in R^w$, and *w* is the width of the window. Also, $\mu x = \overline{X}$ and $\mu y = \overline{Y}$, where $\mu x$ and $\mu y \in R^w$.

In each sequence of the MW, the $\lambda$ is computed and the corresponding values stored in a matrix called Dissimilarity Index Matrix (DIM). $DIM_{(k,l)}$ is a triangular matrix containing the values of $\lambda$, with zeros on its upper diagonal. Each value of $\lambda$ is labeled with the coordinates *k* and *l*, where *k* indicates the position of the spectrum where the windo*w* starts and *l* indicates the *w* used. To investigate the proposed method, we carried out a series of experiments concerning tumour types that, in some cases, had been investigated by other authors and in others, to the best of our knowledge, had not been previously investigated from this point of view.

A visualization of the DIM was obtained for every experiment involving two types of tumours. Figure 2 is an illustrative example, and shows the DIM for *G2[1] vs Mm* and *gl vs me*. This figure shows three distinct zones: Zone 1 (Z1), which corresponds to artifacts associated with the beginnings of both the spectrum and the resonance frequency of water. Zone 2 (Z2) is the spectral band where most metabolic information resides and, therefore, should contribute the most to the discrimination between types of tumours. Zone 3 (Z3), finally, mostly contains noisy, irrelevant information. An analysis of $\lambda$ values and a graphical representation of DIM provide us with interesting insights on the discrimination of tumour types. In brief:

---

[1] G2: high-grade malignant tumours: *gl + me*.

- The highest values of the DIM in small bandwidths (identified as gray belts in figure 2a) correspond to areas of the spectrum that, in previous studies, have been identified as most informative for the differentiation of the corresponding types of tumours. This reinforces the reliability of the proposed method.

- In contrast, in experiments aimed to discriminate between two classes where the percentage of correct classification was reported to stay between 60% and 70% (*gl vs me* and *a2 vs a3*), the values of λ in Z2 fall to levels very close to those obtained for Z1 and Z3 (see Figure 2b), which are known to be completely uninformative (being Z2 the informative area). This shows the potential of the parameter λ as an indicator of the separation between types of tumour, compared with the experiments with high accuracy (e.g., *G2 vs mm*).

*2.3. Detailed Study of DIM and Energy Computation for Feature Selection*

In all experiments, the largest values for λ were found for $w = 1$. Thus, for subsequent experiments, it was decided to use window widths equal to 1. The standard energies (E1, E2 and E3) were calculated for areas Z1, Z2 and Z3, in each experiment. The calculation of energy standardized by areas is given by:

$$E_i = \frac{\sum_{n=1}^{p}\left(\|x_i[n]\|^2\right)}{p}$$

$E_i$ corresponds to the calculation of the standard energy in zone $i$. $x_i[n]$ corresponds to the signal composed of discrete λ values with $w = 1$, of the zone $i$, and $p$ is the size of $x_i[n]$. Given that Z3 does not provide relevant information for classification, due to its direct correspondence with spectral zones modeled as random noise, it was decided to use the energy of this area to rescale the energy of zones $Z_1$ and $Z_2$ by $E_3$.

*2.4. Feature Selection Based on Energy Criteria*

As a preliminary step before the presentation of the patterns to a classifier, an initial feature selection was carried out, obtaining a ranking (in descending order) of the $λ_s$ of the starting window corresponding to zones $Z_1$ and $Z_2$. The variables were divided in groups corresponding to the set of variables whose energy is gradually providing 1% of the total energy of the two zones $E'_i /E_3$, $i=\{1,...\}$ (i.e., $E'_3$ corresponds to a 3% accumulated energy). When the energy of each group is gradually increased, the number of variables also increases, because the number of variables depends strongly of the energy level that everyone provides.

*2.5. Data Classification*

Feed-forward ANN with a 2-layer architecture (i.e. one hidden layer of neurons and one output layer) were used in these experiments for data classification. The input layer of every network was adjusted to the dimension of each set of variable group. Each network had 20 units in the hidden layer and one unit in the output layer. The activation functions *logsig* and *tansig* were used in the hidden layer and the output layer, respectively. The networks were trained with Bayesian regularization back-propagation, which updates the weights and bias according to the algorithm optimization Levenberg-Marquardt [9, 10]. One run of a 5-fold cross-validation was performed for each network, with a maximum of 150 epochs.

As a reference, Table 2 reports the classification results in ascending order ranking of the ratio $E'_{10} / E_3$ with its corresponding classification percentage, and the best classification results reached by others percentage of energy that in most of cases are different of the ratio $E'_{10} / E_3$. Figure 3 shows the logarithmic adjustment of the $\lambda$ values of $E'_{10}/E_3$ versus the mean classification percentage. This figure shows the tendency of the increment of the percentage classification with the increment of $\lambda$.

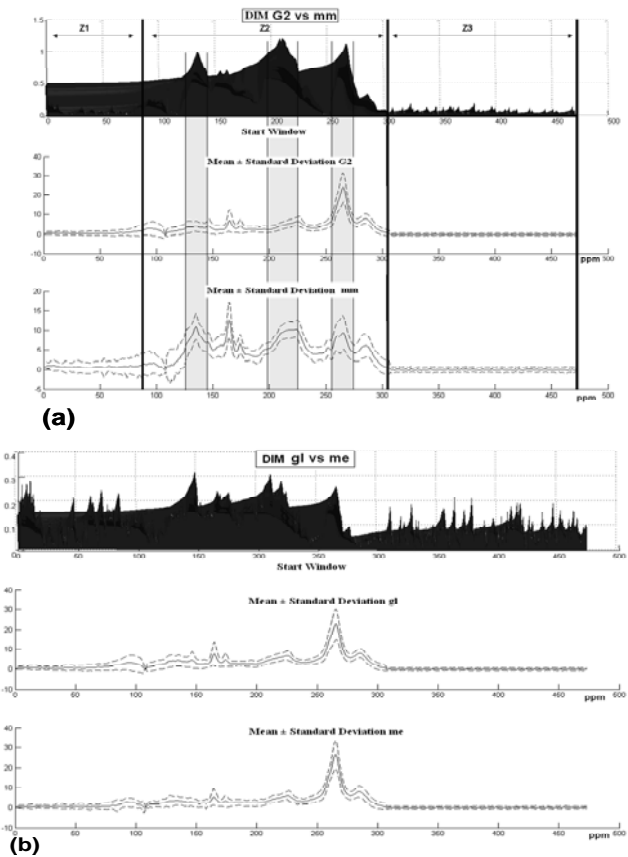

**Figure 2:** Figure 2a (top) is the graphic illustration of DIM zones Z1, Z2 and Z3 for *G2 vs mm*. The gray stripes indicate the areas that generate the biggest differences between *G2* and *mm*, and are shown to correspond to the highest values in the DIM. Figure 2b (bottom) represents the Graphic illustration of DIM for *Gl vs me*, with the representation of the mean (± standard deviation) of the spectra corresponding to each type of tumour. These two types of tumours are particularly difficult to separate, which is reflected in the little difference between the values of the DIM for Z1, Z2 and of Z3.

| Experiments | | | $E'_{10}/E_3$ | Classification percentage | | |
|---|---|---|---|---|---|---|
| | | | | Mean± st. dev. of $E'_{10}/E_3$ | Mean ± st. dev. of maximum values | Energy percentages of maximum values |
| a2 | | a3 | 2.48 | 68.67± 18.5 | 82.00 ± 20.5 | $E'_6/E_3$ |
| gl | | me | 3.36 | 68.38 ± 9.0 | 75.69 ± 9.8 | $E'_8/E_3$ |
| od | | a2 | 3.65 | 68.00 ± 11.0 | 84.00 ± 16.7 | $E'_6/E_3$ |
| a2 | | oa | 3.77 | 76.67 ± 9.4 | 84.67 ± 16.6 | $E'_9/E_3$ |
| gl | | ly | 3.93 | 82.04 ± 12.4 | 88.61 ± 5.5 | $E'_8/E_3$ |
| gl | | ab | 3.97 | 94.18 ± 5.9 | 95.29 ± 6.4 | $E'_9/E_3$ |
| me | | ly | 5.14 | 82.50 ± 16.8 | 82.50 ± 6.8 | $E'_2/E_3$ |
| gl | vs. | a3 | 5.68 | 90.00 ± 8.4 | 94.00 ± 4.5 | $E'_7/E_3$ |
| a2 | | ly | 5.87 | 81.78 ± 20.4 | 89.78 ± 10.0 | $E'_4/E_3$ |
| gl | | pn | 7.85 | 91.76 ± 8.9 | 97.65 ± 3.2 | $E'_5/E_3$ |
| me | | pn | 9.53 | 92.00 ± 8.2 | 93.14 ± 9.6 | $E'_8/E_3$ |
| mm | | ab | 10.25 | 91.00 ± 12.3 | 93.33 ± 8.6 | $E'_6/E_3$ |
| G1* | | mm | 11.42 | 97.67 ± 3.2 | 100 ± 0.0 | $E'_9/E_3$ |
| a2 | | G2 | 15.27 | 93.10 ± 3.7 | 94.51 ± 4.6 | $E'_3/E_3$ |
| G1* | | G2 | 16.35 | 92.10 ± 3.0 | 94.29 ± 5.4 | $E'_7/E_3$ |
| me | | mm | 17.68 | 88.43 ± 9.2 | 98.00 ± 4.5 | $E'_3/E_3$ |
| G2 | | mm | 18.75 | 88.33 ± 2.9 | 89.51 ± 4.9 | $E'_9/E_3$ |
| G1* | | no | 23.88 | 98.89 ± 2.5 | 100 ± 0.0 | $E'_3/E_3$ |
| me | | no | 26.96 | 97.78 ± 5.0 | 98.00 ± 4.5 | $E'_7/E_3$ |

**Table 2**: Ranking (in ascending order) of the ratio $E'_{10}/E_3$ with its corresponding classification percentage. *G1 are low grade gliomas (a2+oa+od).

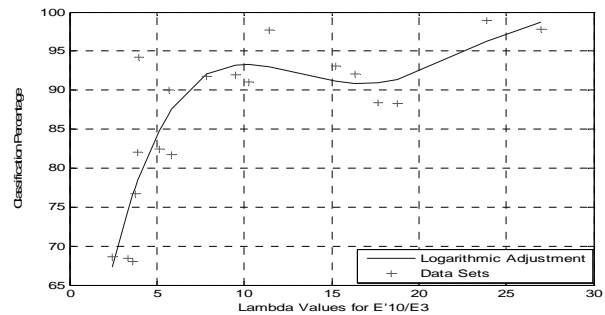

**Figure 3:** Logarithmic adjustment (4$^{th}$ order) of λ values $E'_{10}/E_3$ vs mean classification percentage of $E'_{10}/E_3$.

**Conclusions**

The experimental results reported in this study reinforce the validity of the defined parameter λ as an indicator of the potential of separation between types of tumour. The λ values confirms the lack of informative content of the spectral regions Z1 and Z3, and agrees with the absence in them of any metabolic resonances of interest for the classification of brain tumours, as reported in the literature of the field.

A relationship between the increase in λ value and the percentage of correct classification reached in different experiments has also been found. Thus, the calculation of these ratios can be used as an indicator to find which frequencies or ranges of frequencies have the greatest ability to discriminate between types of tumor, as well as to investigate the degree of overlapping between types of tumour as a preliminary step before the presentation of the patterns to a classifier.

In most cases, the percentage of classification exceeded the values reached by other studies [11-13] using different techniques, being of special relevance the experiments *gl vs me* and *a2 vs a3*, which according to existing literature, are specially difficult classification problems.

**Acknowledgements**

Authors gratefully acknowledge the former INTERPRET (EU-IST-1999-10310) European project partners. Data providers: Dr. C. Majós (IDI), Dr.À.Moreno-Torres (CDP), Dr. F.A. Howe and Prof. J. Griffiths (SGUL), Prof. A. Heerschap (RU), Dr. W. Gajewicz (MUL) and Dr. J. Calvar (FLENI); data curators: Dr. A.P. Candiota, Ms. T. Delgado, Ms. J. Martín, Dr. I. Olier and Mr. A. Pérez (all from GABRMN-UAB). C. Arús and M. Julià-Sapé are funded by the CIBER of Bioengineering, Biomaterials and Nanomedicine, an initiative of the Instituto de Salud Carlos III (ISCIII) of Spain. This work was supported by the Spanish MICINN under grant TIN 2006-08114.